\documentclass[aps,prb,epsfigm,twocolumn,showpacs]{revtex4}
\usepackage{graphicx}
\usepackage{amsmath}

\begin{document}

\title{ Tuning the tunnel coupling of quantum dot molecules with longitudinal magnetic fields }

\author{J. I. Climente}
\email{climente@qfa.uji.es}
\affiliation{Departament de Qu\'{\i}mica F\'{\i}sica i Anal\'{\i}tica,
Universitat Jaume I, Box 224, E-12080, Castell\'o, Spain}
\date{\today}

\begin{abstract}

We show that the energy splitting between the bonding and antibonding 
molecular states of holes in vertically stacked quantum dots 
can be tuned using longitudinal magnetic fields.
With increasing field, the energy splitting first decreases down to zero
and then to negative values, which implies a bonding-to-antibonding ground 
state transition. 
This effect is a consequence of the enhancement of the valence band spin-orbit interaction
induced by the magnetic field, and it provides a flexible mechanism to switch the molecular
ground state from bonding to antibonding. 

\end{abstract}


\maketitle


When two semiconductor quantum dots (QDs) are placed close to each other,
the atomic-like states of the individual dots hybridize forming bonding (nodeless)
and antibonding (noded) molecular-like states, in analogy with diatomic 
molecules.\cite{BryantPRB,HolleitnerSCI,PiPRL,BayerSCI}
The energy splitting between the bonding and antibonding states is given by 
tunnel coupling strength, i.~e.~the overlap between the atomic-like orbitals in 
the interdot barrier. 
The ability to manipulate this coupling in a controllable way while preserving
the quantum coherence is important for the development of various device applications 
of coupled quantum dots (CQDs) in spintronics\cite{WolfSCI}, optoelectronics\cite{Grundmann_book}, 
photovoltaics\cite{NozikPE} and quantum information technologies\cite{BayerSCI,TroianiPRL,StinaffSCI}.

In electrostatically confined CQDs, where the dots are usually laterally coupled, 
accurate control can be achieved through the gate voltage or perpendicular magnetic fields.\cite{HuttelPRB}
In vertically CQDs however the degree of control achieved to date is comparatively lower. 
In these structures, the potential barrier height is fixed by the band-offset 
between the QD and the surrounding matrix materials.
By increasing the barrier length, one can reduce the tunnel coupling and 
hence the splitting between bonding and antibonding levels, $\Delta_{\mbox{{\tiny{BAB}}}}$,\cite{StinaffSCI,KrennerPRL,BrackerAPL}
but this can only be done during the sample growth.
A more flexible method is the use of transverse magnetic fields, which enable to tune 
the tunnel coupling of a sample after its growth.\cite{BurkardPRB,KorkusinskiPRB,BellucciPRB}
Yet, large transverse fields may be required to obtain a sizeable reduction 
of $\Delta_{\mbox{{\tiny{BAB}}}}$ --because the vertical confinement of
these structures is usually strong-- and the axial symmetry of the 
structure is broken by the field. The latter effect activates otherwise forbidden 
transitions that are undesirable for optical manipulation and non-destructive
measurements.\cite{Kim_arxiv}
The use of longitudinal magnetic fields would clearly be more desirable,
but it has been shown that they barely affect the molecular coupling of electrons
except in some particular setups, such as asymmetric coupled quantum rings.\cite{DaSilvaPRB}

In this work we show that an unprecedented degree of control on the tunnel coupling
of vertically CQDs can be achieved with longitudinal magnetic fields if instead of
using conduction electrons one uses valence holes. The method can be applied to
regular self-assembled or litographically grown QDs, and it is particularly
suitable for quantum information devices, where holes may outperform electrons 
due to their reduced spin relaxation and decoherence rates.\cite{GerardotNAT}
The method follows from recent theoretical\cite{ClimentePRB} and experimental\cite{DotyXXX}
findings, which revealed that the valence band mixing of holes was responsible for a 
striking bonding-to-antibonding reversal of the ground state of CQDs with increasing 
interdot distance.
Here we show that the such a reversal can be also induced by a longitudinal magnetic field, 
as it enhances the valence band mixing.

In order to investigate this effect we describe the hole states in CQDs using a 
four-band Luttinger-Kohn k$\cdot$p Hamiltonian, which includes heavy hole (HH) 
and light hole (LH) coupling via spin-orbit (SO) interaction. Details about the
theoretical method are given in Ref.~\onlinecite{ClimentePRB}.
This model correctly described the qualitative features observed in related experiments.\cite{DotyXXX} 
A magnetic field along the $z$ direction is included
using a vector potential in the symmetric gauge, $\mathbf{A} = B/2\,(-y,x,0)$.\cite{RegoPRB} 
The spin Zeeman splitting is neglected, as it simply provides a small numerical 
correction to the orbital effects discussed here.
The QDs we consider are disk-shaped and they have circular symmetry. Thus,
the CQD potential reads $V(\rho,z)=V(\rho)+V(z)$. Here, $V(\rho)$ is an infinite 
well and $V(z)$ is a double rectangular well, whose value is zero inside the dots, $V_{c}$ 
(the band-offset potential) in the interdot region and infinite elsewhere.
The lowest hole states are the Luttinger spinors with total angular momentum
(Bloch + envelope) $z$-component $F_z=+3/2$, and chirality symmetry up ($\nu=\uparrow$)
or down ($\nu=\downarrow$):\cite{ClimentePRB}

\begin{equation}
|F_z=3/2,\nu=\uparrow \rangle = 
\left(
\begin{array}{l}
c_{\scriptscriptstyle{+3/2}}\, f_{0}(\rho,\theta)\,\xi_b(z)\, |J_z=+\frac{3}{2}\rangle \\
c_{\scriptscriptstyle{-1/2}}\, f_{2}(\rho,\theta)\,\xi_b(z)\, |J_z=-\frac{1}{2}\rangle \\
c_{\scriptscriptstyle{+1/2}}\, f_{1}(\rho,\theta)\,\xi_{ab}(z)\, |J_z=+\frac{1}{2}\rangle \\
c_{\scriptscriptstyle{-3/2}}\, f_{3}(\rho,\theta)\,\xi_{ab}(z)\, |J_z=-\frac{3}{2}\rangle \\
\end{array}
\right),
\label{eq1}
\end{equation}

\noindent and

\begin{equation}
|F_z=3/2,\nu=\downarrow \rangle = 
\left(
\begin{array}{l}
c_{\scriptscriptstyle{+3/2}}\, f_{0}(\rho,\theta)\,\xi_{ab}(z)\, |J_z=+\frac{3}{2}\rangle \\
c_{\scriptscriptstyle{-1/2}}\, f_{2}(\rho,\theta)\,\xi_{ab}(z)\, |J_z=-\frac{1}{2}\rangle \\
c_{\scriptscriptstyle{+1/2}}\, f_{1}(\rho,\theta)\,\xi_{b}(z)\, |J_z=+\frac{1}{2}\rangle \\
c_{\scriptscriptstyle{-3/2}}\, f_{3}(\rho,\theta)\,\xi_{b}(z)\, |J_z=-\frac{3}{2}\rangle \\
\end{array}
\right).
\label{eq2}
\end{equation}

\noindent Here $f_{m_z} (\rho,\theta)$ represents the in-plane part of the envelope function, 
with envelope angular momentum $m_z$. $\xi_{\nu_z}(z)$ is the vertical part of the envelope
function, which can be bonding ($\nu_z=b$) or antibonding ($\nu_z=ab$).\cite{bab} 
$|J_z\rangle$ represents the Bloch function with Bloch angular momentum $J_z$, and
$c_{\scriptscriptstyle{J_z}}$ is a coefficient that we determine numerically.\cite{basis}
In the absence of magnetic fields, $|F_z=3/2,\,\nu=\uparrow\rangle$ is Kramers-degenerate 
with $|F_z=-3/2,\,\nu=\downarrow\rangle$,  and $|F_z=3/2,\,\nu=\downarrow\rangle$ with 
$|F_z=-3/2,\,\nu=\downarrow \rangle$. However, the field lifts this degeneracy favoring
the states with positive $F_z$,\cite{RegoPRB} so that Eqs.~(\ref{eq1}) and (\ref{eq2}) soon
describe the two lowest-lying levels. We shall focus on these states. 

For a usual CQD, the first component of the spinors above is strongly dominant.
Thus, $|F_z=3/2,\, \nu=\uparrow \rangle$ is essentially a bonding HH with $m_z=0$ 
and $|F_z=3/2, \nu=\downarrow \rangle$ is essentially an antibonding HH with $m_z=0$.
Yet, the minor LH component with $m_z=1$ ($J_z=+1/2$) can bring about important 
changes in the molecular behavior.
This is because its molecular character (bonding or antibonding) is opposite to that 
of the dominant component. In the spinor given by Eq.~(\ref{eq1}), the LH component
is antibonding, so that it unstabilizes the bonding HH. By contrast, in the spinor of
Eq.~(\ref{eq2}) it is bonding and therefore stabilizes the antibonding HH.
This results in an overall reduction of $\Delta_{\mbox{{\tiny{BAB}}}}$,
and we say that the tunnel coupling strength has been reduced by the SO induced
valence mixing.\cite{ClimentePRB}
Since the tunneling of LHs is much larger than that of HHs, this effect 
can be important even if the weight of the LH component is small. 
Indeed, for long interdot distances, where the tunneling of HHs is neglegible compared to
that of LHs, this is responsible for the reversal of $|F_z=3/2,\, \nu=\uparrow \rangle$ 
and $|F_z=3/2,\,\nu = \downarrow \rangle$ spinors observed by Doty et al.\cite{DotyXXX}
The main message of this paper is that a similar control on $\Delta_{\mbox{{\tiny{BAB}}}}$
can be achieved not by changing the interdot distance, but simply by applying a 
longitudinal magnetic field. 
This is because the dominant HH component has $m_z=0$, but the relevant LH component 
has $m_z=1$. 
Thus, while the HH component is little sensitive to the field, the LH component is 
stabilized, its relative weight increasing gradually with $B$.

\begin{figure}[h]
\includegraphics[width=0.4\textwidth]{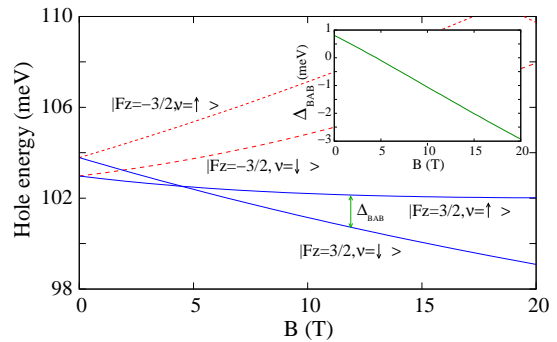}
\caption{(Color online). Hole energy levels in a CQD as a function of the magnetic field.
The inset shows the splitting between the lowest ``bonding'' and ``antibonding'' levels, 
which is tuned by the field down to zero and then to negative values.}\label{Fig1} 
\end{figure}

To illustrate this principle, in Fig.~\ref{Fig1} we plot the lowest hole levels of a 
GaAs/Al$_{0.3}$Ga$_{0.7}$As CQD vs.~magnetic field. 
The two CQDs are identical and form a homonuclear QD molecule.\cite{datafig1} 
Solid lines are used for states with positive $F_z$ and dashed ones
for states with negative $F_z$. When the field is switched on $|F_z=3/2,\, \nu=\uparrow \rangle$ 
becomes the ground state, but for stronger fields ($B > 4.5$ T) it is replaced by 
$|F_z=3/2,\, \nu=\downarrow \rangle$. This is because $|F_z=3/2,\,\nu=\downarrow\rangle$
has a larger LH component than $|F_z=3/2,\,\nu=\uparrow\rangle$,\cite{ClimentePRB}
and is then more strongly affected by the field.
One can identify the energy splitting between these two states with $\Delta_{\mbox{{\tiny{BAB}}}}$, 
which we plot in the inset (solid line). As can be seen, the magnetic field tunes
$\Delta_{\mbox{{\tiny{BAB}}}}$ from initially positive values down to zero and then to negative values.
Actually, the negative values induced at large magnetic fields can be even larger (in magnitude)
than those at zero field.
This shows that the longitudinal magnetic field is a versatile tool to manipulate the 
tunnel coupling strength.
We note that $\Delta_{\mbox{{\tiny{BAB}}}}$ can be tuned to zero because $|F_z=3/2,\, \nu=\uparrow \rangle$ 
and $|F_z=3/2,\, \nu=\downarrow \rangle$ have different chirality symmetry, which is preserved by the field.
If this was not the case, the two states would anticross and $\Delta_{\mbox{{\tiny{BAB}}}}$ would 
never be zero.

For the magnetic-field-induced bonding-antibonding reversal to take place,
a number of conditions must be met. First, the ground state at zero field must be
$|F_z=3/2,\, \nu=\uparrow \rangle$, which is usually the case for small interdot
distances.\cite{ClimentePRB,DotyXXX} Otherwise the ground state will always be 
$|F_z=3/2,\, \nu=\downarrow \rangle$, as its splitting with $B$ is larger.
Second, the LH component must be sizeable. This depends on the constituent
materials and the strain fields.\cite{ClimentePRB} We have tested that the 
ground state reversal of Fig.~\ref{Fig1} is also feasible in InGaAs/GaAs CQDs, 
but it takes place at much larger values of $B$ (not shown), because the 
biaxial strain severely weakens the HH-LH mixing.\cite{KorkusinskiPRB}

\begin{figure}[h]
\includegraphics[width=0.4\textwidth]{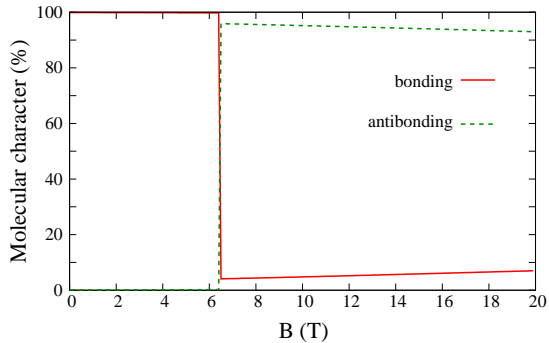}
\caption{(Color online). Molecular character of the ground state as a function of the magnetic field.
Solid line indicates the weight of bonding character, while dashed line indicates that
of antibonding character. Note the abrupt transition, from mostly bonding to mostly
antibonding, at $B \sim 4.5$ T.}\label{Fig2}
\end{figure}

Since Luttinger spinors contain an admixture of bonding and antibonding components, 
it is worth quantifying to which extent the ground state reversal reported in Fig.~\ref{Fig1} 
implies a change in the molecular character. To this end, in Fig.~\ref{Fig2} we plot the weight 
of the ground state spinor components with bonding (solid line) and antibonding (dashed line) 
character as a function of the magnetic field, as inferred from the squared coefficients of 
Eqs.~(\ref{eq1}) and (\ref{eq2}). 
Up to $B \sim 4.5$ T, when the ground state is $|F_z=3/2,\, \nu=\uparrow \rangle$,
the molecular character is over 99\% bonding.  For stronger fields, with 
$|F_z=3/2,\, \nu=\downarrow \rangle$ as the ground state, it is 93-96 \% antibonding. 
This clearly confirms that the ground state is switched from mostly bonding to mostly antibonding.
The stronger admixture of $|F_z=3/2,\, \nu=\downarrow \rangle$ is connected with its
larger LH component, as mentioned above.

So far we have investigated the simple case of a single hole in two identical
QDs. A more realistic scenario, which could serve to test our predictions, is 
studied next. We consider two dots with slightly different size and charged with an interacting
electron-hole pair (exciton), as in usual optically-charged self-assembled CQDs. 
In this kind of systems the particles are mostly localized in the bigger QD, so
that the molecule is strongly heteronuclear. However, one can apply an electric field
along the molecular axis to induce a resonance of the atomic-like levels of either
electrons or holes.\cite{BrackerAPL}
In this way, a homonuclear-like behavior is restored for the chosen carrier.
At the resonant value of the electric field, the emission spectrum of the
exciton reveals an anticrossing between the bonding and antibonding states
whose magnitude is precisely $\Delta_{\mbox{\tiny BAB}}$.\cite{KrennerPRL,BrackerAPL,OrtnerPRL}

\begin{figure}[h]
\includegraphics[width=0.4\textwidth]{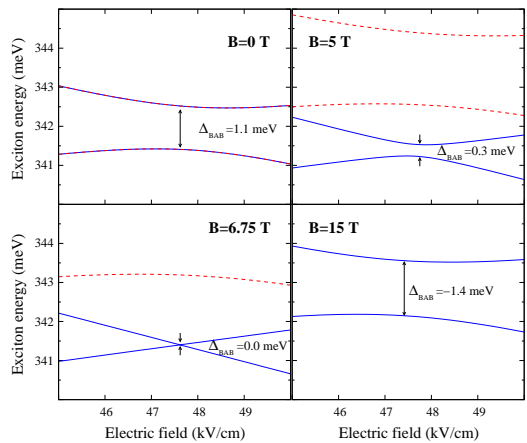}
\caption{(Color online). Exciton energy as a function of the electric field for an asymmetric
CQD subject to different magnetic fields. Solid lines denote the states involving
hole levels with $F_z=3/2$, while dashed lines denote those involving $F_z=-3/2$.
The gap between the ``bonding'' and ``antibonding'' exciton levels is 
tuned with the magnetic field.}\label{Fig3}
\end{figure}

In Fig.~\ref{Fig3} we plot the low-energy exciton states as a function of the
magnetic field for an asymmetric CQD.\cite{datafig3} The electron state is calculated
using an effective mass approach and the electron-hole Coulomb interaction term is
solved using a configuration interaction scheme with the Hartree products arising
from the two lowest (bonding and antibonding) electron and hole states.\cite{ClimentePRB} 
An electric field is applied which brings the hole into resonance while leaving
the electron in the higher dot.
The fine structure arising from electron-hole exchange interaction is
neglected here as it is not relevant for the message.
Four different cases are illustrated. At zero magnetic field (top left panel), 
an anticrossing of $\Delta_{\mbox{\tiny BAB}}=1.1$ meV is observed.
At $B=5$ T (top right panel), this gap is reduced to $\Delta_{\mbox{\tiny BAB}}=0.3$ meV,
as the hole states $|F_z=3/2,\,\nu=\uparrow\rangle$ and $|F_z=3/2,\,\nu=\downarrow\rangle$
are now closer together.
At $B=6.75$ T (bottom left panel), the gap collapses, indicating that the two hole 
levels are degenerate.
Note that this occurs in spite of the fact that the different vertical
confinement of the top and bottom dots breaks the chirality symmetry.
The reason for this is that the resonant electric field restores an 
effective chirality for the molecular states.\cite{ClimentePRB,DotyXXX}
Finally, at $B=15$ T (bottom right panel) $|F_z=3/2,\,\nu=\downarrow\rangle$
is by far the hole ground state, and the anticrossing gap ($\Delta_{\mbox{\tiny BAB}}=-1.4$ meV)
is even larger in magnitude than that it was at zero field. Further, it has 
negative sign because the ground state is mostly antibonding.
Similar control of the anticrossing gaps may be expected in more complex
species as long as holes are the resonant carrier, because the gap is ultimately
dependent on the hole tunnel coupling strength.\cite{ScheibnerPRB}

In conclusion, we have demonstrated that the tunnel coupling of QD
molecules containing resonant holes can be controlled using longitudinal magnetic 
fields. The tunnel coupling strength can be reduced down to zero and to large
negative values, thus switching from a ground state with strong 
bonding character to one with strong antibonding character.
This tuning of the molecular spectrum is exclusive of artificial
molecules, due to the SO-induced valence band mixing, and 
it can be exploited to produce quantitative or qualitative changes in the
response of devices based on vertically CQDs. 
The tunneling strength of resonant holes can be accurately measured by
single QD molecule photoluminescence\cite{BrackerAPL}, and the bonding
or antibonding character of the ground state by 
magneto-photoluminescence.\cite{DotyXXX}
We then propose experiments to verify our predictions.

We thank P. Hawrylak, M. Doty and D. Gammon for critical reading of
the manuscript. Support from the Ramon y Cajal program, MEC project 
CTQ2008-03344 and Cineca Calcolo Parallelo 2008 is acknowledged.


\newpage

\begin{thebibliography}{30}

\bibitem{BryantPRB}
G.W. Bryant, Phys. Rev. B {\bf 47}, 1683 (1993).

\bibitem{HolleitnerSCI}
A.W. Holleitner, R.H. Blick, A.K. H\"uttel, K. Eberl, J.P. Kotthaus,
Science {\bf 297}, 5578 (2002).

\bibitem{PiPRL}
M. Pi, A. Emperador, M. Barranco, F. Garcias, K. Muraki, S. Tarucha, and D. G. Austing,
Phys. Rev. Lett. {\bf 87}, 066801 (2001).

\bibitem{BayerSCI}
M. Bayer, P. Hawrylak, K. Hinzer, S. Fafard, M. Korkusinski,
Z.R. Wasilewski, O. Stern, and A. Forchel, Science {\bf 291}, 451 (2001).

\bibitem{WolfSCI}
S. A. Wolf, D. D. Awschalom, R. A. Buhrman, J. M. Daughton, S. von Moln\'ar, 
M. L. Roukes, A. Y. Chtchelkanova, and D. M. Treger,
Science {\bf 294}, 1488 (2001).

\bibitem{Grundmann_book}
M. Grundmann (Ed.),  \emph{Nano-optoelectronics: Concepts, Physics and Devices},
(Springer-Verlag, Berlin, 2002).

\bibitem{NozikPE}
A.J. Nozik, Physica E (Amsterdam) {\bf 14}, 115 (2002).

\bibitem{TroianiPRL}
F. Troiani, E. Molinari, and U. Hohenester,
Phys. Rev. Lett. {\bf 90}, 206802 (2003).

\bibitem{StinaffSCI}
E.A. Stinaff, M. Schneibner, A.S. Bracker, I.V. Ponomarev, V.L. Korenev,
M.E. Ware, M.F. Doty, T.L. Reinecke, and D. Gammon, Science {\bf 311}, 636 (2006).

\bibitem{HuttelPRB}
A.K. H\"uttel, S. Ludwig, H. Lorenz, K. Eberl, and J.P. Kotthaus,
Phys. Rev. B {\bf 72}, 081310(R) (2005).

\bibitem{KrennerPRL}
H.J. Krenner, M. Sabathil, E.C. Clark, A. Kress, D. Schuh, M. Bichler,
G. Abstreiter, and J.J. Finley, Phys. Rev. Lett. {\bf 94}, 057402 (2005);

\bibitem{BrackerAPL}
A.S. Bracker, M. Schneiber, M.F. Doty, E.A. Stinaff, I.V. Ponomarev,
J.C. Kim, L.J. Whitman, T.L. Reinecke, and D. Gammon, Appl. Phys. Lett. 
{\bf 89}, 233110 (2006).

\bibitem{BurkardPRB}
G. Burkard, G. Seeling, and D. Loss,
Phys. Rev. B {\bf 62}, 2581 (2000).

\bibitem{KorkusinskiPRB}
M. Korkusinski, and P. Hawrylak, Phys. Rev. B {\bf 63}, 195311 (2001).

\bibitem{BellucciPRB}
D. Bellucci, F. Troiani, G. Goldoni, and E. Molinari,
Phys. Rev. B {\bf 70}, 205332 (2004).

\bibitem{Kim_arxiv}
D. Kim, S.E. Economou, S.C. Badescu, M. Scheibner, A.S. Bracker, M. Bashkansky,
T.L. Reinecke, and D. Gammon, arxiv:0809.1673v1.

\bibitem{DaSilvaPRB}
L.G.G.V. Dias da Silva, J.M. Villas-Boas, and S. Ulloa,
Phys. Rev. B {\bf 76}, 155306 (2007).

\bibitem{GerardotNAT}
B.D. Gerardot, D. Brunner, P.A. Dalgarno, P. \"Ohberg, S. Seidl, M. Kroner,
K. Karrai, N.G. Stoltz, P.M. Petroff, and R. Warburton,
Nature (London) {\bf 451}, 441.

\bibitem{ClimentePRB}
J.I. Climente, M. Korkusinski, G. Goldoni, and P. Hawrylak,
Phys. Rev. B {\bf 78}, 115323 (2008).

\bibitem{DotyXXX}
M.F. Doty, J.I. Climente, M. Korkusinski, M. Scheibner, A.S. Bracker, P. Hawrylak, and D. Gammon,
submitted (arXiv:0804.3097v1).

\bibitem{RegoPRB}
L.G.C. Rego, P. Hawrylak, J.A. Brum, and A. Wojs, Phys. Rev. B {\bf 55}, 15694 (1997).

\bibitem{bab} When the two QDs of the molecule are identical, bonding and antibonding are 
simply the symmetric and antisymmetric linear combinations of atomic orbitals.

\bibitem{datafig1} The QDs are $2.5$ nm high and have a radius of 15 nm. The interdot 
barrier is $3.5$ nm long with height $V_c=200$ meV. GaAs Luttinger parameters are used, 
$\gamma_1=6.98$, $\gamma_2=2.06$, $\gamma_3=2.93$.\cite{VurgaftmanJAP}

\bibitem{VurgaftmanJAP}
I. Vurgaftman, J.R. Meyer, L.R. Ram-Mohan, J. Appl. Phys. {\bf 89}, 5815 (2001).

\bibitem{basis}
For holes (electrons), $f_{m_z} (\rho,\theta)$ is expanded on a basis of 6 (3)
Bessel functions and $\xi_{\nu_z}(z)$ on a basis of 30 harmonics. The coefficients
are found with an exact diagonalization procedure. 

\bibitem{OrtnerPRL}
G. Ortner, M. Bayer, Y. Lyanda-Geller, T.L. Reinecke, A. Kress,
J.P. Reithmaier, and A. Forchel, Phys. Rev. Lett. {\bf 94}, 157401 (2005).

\bibitem{datafig3}
The upper (lower) QD is now 2.7 (2.3) nm high.
The electron effective mass is $m_e^*=0.067\,m_0$ ($m_0$ is the free electron mass),
 the conduction barrier potential is $V_c^e=260$ meV and the dielectric constant
is $\epsilon=12.4$. The rest of parameters are as in Ref.\onlinecite{datafig1}.

\bibitem{ScheibnerPRB}
M. Scheibner, M.F. Doty, I.V. Ponomarev, A.S. Bracker, E.A. Stinaff,
V.L. Korenev, T.L. Reinecke, and D. Gammon, 
Phys. Rev. B {\bf 75}, 245318 (2007).

\end{thebibliography}
\end{document}